# Biased quasi ballistic spin torque magnetization reversal

*S. Serrano-Guisan[1], K. Rott[2], G. Reiss[2], J. Langer[3], B. Ocker[3] and H. W. Schumacher[1]*

1) Physikalisch-Technische Bundesanstalt, Bundesallee 100, D-38116 Braunschweig, Germany.

2) University of Bielefeld Department of Physics Universitätsstr. 25, 33615 Bielefeld Germany.

3) Singulus Nano Deposition Technologies GmbH, Hanauer Landstrasse 103, D-63796 Kahl am Main, Germany

Corresponding authors:

E-mail:      santiago.serrano-guisan@ptb.de
phone:       +49 (0)531 592 2439,      fax:    +49 (0)531 592 2205

E-mail:      hans.w.schumacher@ptb.de
phone:       +49 (0)531 592 2414,      fax:    +49 (0)531 592 69 2414

## Abstract

We explore the fundamental time limit of ultra fast spin torque induced magnetization reversal of a magnetic memory cell. Spin torque precession during a spin torque current pulse and free precessional magnetization ringing after spin torque pulse excitation is detected by time resolved magneto transport. Adapting the duration of the spin torque excitation pulse to the spin torque precession period allows suppression of the magnetization ringing and thus coherent control of the final orientation of the magnetization. In the presence of a hard axis bias field such coherent control enables an optimum ultra fast, quasi ballistic spin torque magnetization reversal by a single precessional turn directly from the initial to the reversed equilibrium state.

# Article

A spin polarized current transfers spin angular momentum to the magnetization M of a ferromagnet and thereby exerts a so-called spin torque (ST) on M (1,2). ST can result in the excitation of spin waves or steady state precession of M (3,4), and in current induced magnetization reversal (5,6,7) enabling a new class of spintronic devices driven by currents instead of magnetic fields. Promising applications are ST oscillators and current programmable non volatile magnetic random access memories (MRAM) (8). ST precession has been intensively studied in frequency domain (9,10,11,12,13,14,15,16); additional time domain measurements (17,18,19,20,21,22,23) have underlined the importance of ST precession for magnetization reversal and showed that M generally undergoes *multiple* precessional turns during reversal (19-23). In contrast ultra fast *field induced* magnetization reversal can occur on the time scale of a *half* precessional turn, only (24,25,26). During this so called *ballistic* magnetization reversal (25,27,28) the duration of a transverse field pulse matches one half of the precession period leading to an optimized reversal trajectory directly from the initial to the reversed equilibrium state during pulse application. Realization of corresponding *ballistic ST reversal* could pave the way for future ultra fast and yet efficient ST devices. Here, we explore the ultra fast limit of ST induced magnetization reversal. We detect ST precession and ST induced free precession of M by time resolved magneto resistance measurments. Adapting the duration of the current pulse to the ST precession period allows a coherent control of the free ringing and thus of the final orientation of M upon current pulse decay. We show that under the presence of a hard axis bias field ultra fast *quasi ballistic* ST reversal of M directly form the initial to the final equilibrium orientation can occur by a single precessional turn. Applications in ultra fast ST memory and logic devices can be foreseen.

Our experiments are performed on nanopillars of MgO based low resistive magnetic tunnelling junctions (MTJ) with a resistance area product of about 6.5 $\Omega\mu m^2$. MTJs are sputter deposited on oxidized Si wafers in ultra high vacuum in a Singulus NDT Timaris cluster tool. The deposition sequence is Ta 3nm / CuN 90nm / Ta 5nm / PtMn 20nm/ $Co_{60}Fe_{20}B_{20}$ 2nm/ Ru 0.75nm/ $Co_{60}Fe_{20}B_{20}$ 2nm/ MgO 1.3nm / $Co_{60}Fe_{20}B_{20}$ 3nm/ Ta 10nm / Cu 30nm/ Ru 8nm. The MgO barrier is formed using a three step sequence. First 0.9nm of metallic Mg are deposited which is then allowed to oxidize in 0.01 Torr oxygen pressure for 300s . In a third step 0.4nm of metallic Mg is deposited on top of the MgO. After deposition the stack is annealed for 90 minutes at 360° C in 1 T field. The stack is then patterned into elliptic nanopillars of 140 nm x 285 nm lateral dimensions by electron beam lithography and ion milling.

Fig. 1(a) shows a room temperature tunnelling magneto resistance (TMR) loop with the field swept along the easy (long) axis of the ellipsoid (x-axis). The magnetization M of the upper 3 nm thick $Co_{60}Fe_{20}B_{20}$ free layer (blue arrow) reverses while the magnetization $M_P$ of the lower pinned layer (red arrow) remains aligned along the negative x-axis. The loop shows well-defined reversal from the parallel low resistance state at negative fields to the high resistance antiparallel state at positive fields with a TMR ratio of 140 %. Fig. 1(b) shows a switching asteroid compiled from x-axis hysteresis loops taken at various in plane hard axis (y-axis) fields. The sample shows well defined quasi single domain reversal behaviour with a uniaxial anisotropy field of $\mu_0 H_K \approx 15$ mT. In static ST measurements such devices show a critical current density for ST magnetization reversal of about $j_C = 7...9 \times 10^6 A/cm^2$.

The ST dynamics of the free M are detected at room temperature by time resolved magneto resistance measurements in reflection as sketched in Fig. 1(c). The MTJ nanopillars are contacted by microwave probes and terminate a 50 $\Omega$ coaxial line. A current pulse from a pulse generator is split by a power splitter and one half of the pulse is injected into the MTJ while the other half is terminated by a 18 GHz sampling oscilloscope. Due to impedance mismatch of the MTJ the current pulse is partially reflected by the MTJ. The reflected pulse is again split by the power splitter and half of the reflected signal is recorded by the sampling oscilloscope while the other half is terminated in the pulse generator. An ST excitation of M by the injected current pulse leads to an angular excursion of M and thus a change of the TMR can be detected in the reflected signal. To separate the TMR change from the background of the reflected pulse a weak DC detection current ($I_{DC}$ = 200 µA) is injected into the MTJ through a bias tee (29). Subtraction of two oscilloscope traces measured at $\pm I_{DC}$ yields the TMR signal change without the background of the reflected pulse. The use of the DC detection current allows both the detection of ST precession *during* application of a ST current pulse as well as the detection of *free* precession (ringing) of M *after* ST pulse application. Therefore, in contrast to earlier studies (19,21,22,23), also *non equilibrium* orientations of M after ST pulse decay can be evidenced and investigated. For the time resolved TMR measurements the oscilloscope traces are taken at 100 kHz pulse repetition rate. Averaging over up to 2000 traces is used to reduce noise level. Two external coils allow application of static fields up to $\mu_0 H_S$ = 30 mT in arbitrary in plane angles $\phi$ as sketched in Fig. 1(b). Current pulses with pulse duration from 160 ps to 5 ns are injected into the MTJ with effective current densities up to j = $9 \times 10^6$ A/cm². Note, that in a reflection setup the superposition of the injected and the reflected pulse at the MTJ leads to an enhancement of the effective pulse current load by a factor of 1 + ($R_{MTJ}$-Z)/($R_{MTJ}$+Z) with $R_{MTJ}$ the junction resistance and Z = 50 Ω the line impedance (30).

Fig. 2 shows two typical measurements of time-resolved ST precession *during* the pulse and free precession of M *after* the pulse of the device characterized in Fig. 1. The TMR voltage change is plotted as a function of time. Fig. 2(a) shows the voltage change during the rise of a 5 ns current pulse of j = $3.2 \times 10^6$ A/cm². The positive j corresponds to spin polarized electrons flowing from the pinned to the free layer thereby favouring the low resistance parallel state. M is on the low resistance side of the asteroid and the easy axis component $m_X < 0$ is oriented parallel to $M_P$ as $\mu_0 H_S = 22.8$ mT $> H_K$ is applied along $\phi = -114°$ (inset sketch). The data clearly shows a damped voltage oscillation on top of a rising background. The oscillation is due to ST precession of M while the background mainly results from an effective tilt $\Delta\phi$ of M towards $M_P$ due to ST. Fig. 2 (c) shows the time resolved TMR voltage change during and after ST excitation by a 180 ps pulse (measured at full width half maximum) of $j = 9 \times 10^6$ A/cm$^2$ with $\mu_0 H_S = 22.8$ and $\phi = 84°$ (see inset sketch). Here, a voltage oscillation *after* the decay of the pulse is clearly observed. Now M relaxes by free precession (ringing) towards its equilibrium orientation along $H_S$.

The response of M during and after ST excitation is usually described by the Landau-Lifshitz-Gilbert (LLG) equation (31) including the Slonczewski ST term (1)

$$\frac{d\vec{M}}{dt} = -\gamma\left(\vec{M} \times \vec{H}_{eff}\right) + \frac{\alpha}{M_s}\left(\vec{M} \times \frac{d\vec{M}}{dt}\right) + a_I \cdot \vec{M} \times \left(\vec{M}_p \times \vec{M}\right).$$

Here, $\gamma$ is the gyromagnetic ratio, $H_{eff}$ is the effective field, i.e. the sum of external and internal (anisotropy, demagnetization) fields, $\alpha$ and $M_S$ are the free layer's Gilbert damping parameter and saturation magnetization, respectively, and $a_I$ is the ST parameter comprising j and the spin polarization of the magnetic layers.

The data of Fig. 2(a),(c) is further analyzed by subtracting an exponential background (red line) resulting in the precession data shown on the right in Fig. 2(b),(d). Fitting the precession data by an exponentially damped sinusoid (red line) yields the precession frequency $f$ and the effective damping $\alpha_{eff}$ (27). The frequencies of free and ST precession $f_{free} = 4.05$ GHz and $f_{ST} = 5.25$ GHz strongly differ. This is mainly due to the different orientations $\phi$ of $H_S$ and thus of M in the two cases. For low applied fields ($H_S$ of the order of $H_K$) the anisotropy of the free layer gives rise to a strong angular dependence of $f$. $f$ is minimum at $\pm$ 90° and maximum along the easy axis leading to a significantly lower precession frequency for the ringing at $\phi = 84°$. $\alpha_{eff}$ is calculated from the decay (27) taking into account $\mu_0 M_S = $ 1.6 T of the free layer yielding $\alpha_{eff}^{ST} = 0.0136 \pm 0.001$ and $\alpha_{eff}^{free} = 0.011 \pm 0.002$. *Free* precession of M is described by the first two terms on the right hand side of the LLG ($a_I = 0$) and $\alpha_{eff}^{free}$ directly yields $\alpha$ of the LLG. In contrast, $\alpha_{eff}^{ST}$ results from competition of Gilbert damping and ST. Here an *increased* effective damping ($\alpha_{eff}^{ST} > \alpha$) is found as expected for the low resistance configuration ($m_X < 0$) with j favouring the low resistance (parallel) state (32,33).

As mentioned above, the free precessional ringing results from a non equilibrium orientation of M at the moment the ST pulse decays. For ultrafast ST-MRAM such non equilibrium orientation of M after a programming pulse is not desirable: First, the precessional relaxation of M to equilibrium can take several ns. As a consequence the *effective* bit programming time from the initial to the final equilibrium state is considerably longer than the ST pulse duration thereby limiting ST-MRAM write performance. Second, high excitation of M can evoke uncontrolled thermally activated reversal of M back to the initial state thereby affecting the switching reliability.

Fig. 3(a) shows that the free ringing of M after application of a ST current pulse can be suppressed by proper selection of the pulse parameters. The time-resolved TMR signal is plotted for six different pulse durations $T_{Pulse}$ from 450 to 1250 ps. The data is taken at $H_S = $ 24 mT, $\phi = -104°$ ($m_X < 0$, low resistance state) with j = $-7.8 \times 10^6$ A/cm$^2$ now favouring the high resistance (antiparallel) state. The curves are offset for clarity. The vertical and the tilted gray line mark the onset and the decay of the excitation pulse, respectively. During pulse application a strong ST precession of $f_{ST} = 3.3$ GHz is observed. Note, that now $\alpha_{eff}^{ST} = 0.0098$

± 0.001 < α for the inverted current polarity favouring the high resistance state (1,2,32,33). A significant ringing after ST pulse decay is observed for $T_{Pulse}$ = 450ps, 770ps, 1100ps (red curves) whereas at $T_{Pulse}$ = 600ps, 930ps and 1250 ps practically no ringing is present after pulse termination (black curves). For the latter the ringing and thus the final orientation of M upon ST pulse decay are coherently controlled. Here, the ST pulse duration $T_{Pulse}$ approximately equals an integer multiple of the ST precession period $T_{Pulse} \approx nT_{Prec}$. During the pulse M undergoes *n* precessional turns starting from $H_S$ and is again aligned along $H_S$ upon pulse decay (lower sketch in (b)) and the ringing is suppressed. For the red curves the ST pulse terminates after approximately n-1/2 precessional turns. At this point of the ST precession trajectory, M has the maximum tilt off equilibrium as shown in the upper sketch in (b). M thus relaxes towards $H_S$ after the pulse by free precessional ringing.

Such coherent control of the final orientation of M after ST excitation can also be used to realize ultra fast ST magnetization reversal. In the following we present time resolved measurements of ST reversal in the presence of a bias field $\mu_0H_S$ = 10 mT along the in plane hard axis (y-axis). Biased ST switching has already been proposed to improve the switching reliability for sub ns ST reversal (22,34). The magnetic configurations are sketched in Fig. 3(d). The experiments are carried out on another MTJ device of similar dimensions but having a higher uniaxial anisotropy of $\mu_0H_K$ = 21.5 mT. Reversal processes from antiparallel to parallel orientation of M by a ST pulse of j > 0 are studied. The initial equilibrium state $M_i$ has a positive easy axis component ($m_X$ > 0) and $H_S$ tilts M to approximately $\phi$ = 60°. The final state $M_f$ is symmetric to the hard axis ($m_{Y,i} = m_{Y,f}$ ; $m_{X,i}$ = - $m_{X,f}$). For the sampling measurements M is reset into the initial antiparallel state by a 10 μs ST reset pulse of j ≈ - $7 \times 10^6$ A/cm² prior to application of the ST switching measurement pulse. Fig. 3(c) shows a time resolved measurement of ST switching by a 5 ns pulse of j = $2.5 \times 10^6$ A/cm². The observed TMR voltage change can be explained by the sketch shown in Fig. 3 (d). At the pulse inset, a strong precessional overshoot is observed followed by a damped precession of M (arrow (i)). Here M overcomes the hard axis and precesses about the intermediate effective field $H_{int}$ resulting from competition of $H_S$, $H_K$ and ST. As ST favours the parallel low resistance state this intermediate orientation is tilted stronger towards $M_P$ than the final reversed orientation $M_f$. After decay of the ST pulse M relaxes from $H_{int}$ to $M_f$ (ii) by free precession. Note however, that precessional oscillations are not resolved but appear as an exponential decay to the final voltage value. Here the precession is dephased due to thermal activation of M after the 5 ns high current density pulse and the oscillations are smeared out in the sampling measurements. Note that the applied bias and the effect of the strong initial precession already allows a highly efficient ST reversal by a current density which is about three times smaller than $j_C = 7...9 \times 10^6 A/cm^2$ derived from static reversal experiments.

Fig. 3 (e) and (f) show reversal by shorter pulses of the same current density $j = 2.5 \times 10^6$ A/cm$^2$. These two pulses decay to zero after 730 ps and 1.1 ns, respectively. For the 730 ps pulse (e) the ST precession terminates during the initial precessional overshoot marked by the arrow ((e) in Fig. 3 (c),(d)). In the data the moment of pulse decay can be observed as a kink in the decay of the overshoot (black arrow). At this moment M has reversed and is oriented between $H_{int}$ and $M_f$. After pulse decay a clear precessional ringing is observed (iii) as M relaxes towards $M_f$ by free precession as sketched in the inset (e). Therefore, despite of the sub ns switching pulse duration the *effective* magnetization reversal time from the initial to the final equilibrium orientation takes several ns. In addition to the ringing, the measured final voltage corresponding to the reversed state is almost a factor of 2 lower than in (c), pointing out to a reduced switching reliability. At the moment of pulse decay M is strongly excited. This excitation can lead to a thermally activated stochastic back switching of M to $M_i$. As the sampling measurements average over the successful and the unsuccessful switching events we deduce that magnetization reversal is only achieved for about 50 % of the applied pulses.

Note that the measured ringing after pulse decay (iii) could in principle also be a consequence of precessional relaxation towards $M_i$ after the unsuccessful switching events. In the case of an unsuccessful switching, M stochastically overcomes the hard axis barrier towards $M_i$ involving thermal activation. Such thermal activation however evokes a considerable dephasing of the precession which should smear out the precessional oscillations from back switching in the sampling measurements.

Fig. 3 (f) shows magnetization reversal by a 1.1 ns ST pulse. Now the ST pulse terminates in the voltage minimum after the precessional overshoot (arrow (f) in Fig. 3 (c),(d)). Again the precessional overshoot is clearly observed. The pulse decays at the same time as the overshoot (arrow) and the TMR voltage immediately equals the final value of the reversed state in (c) evidencing a highly reliable magnetization reversal. No significant free precessional ringing of M after the overshoot is found and no exponential decay speaking for a dephased precession is present. Here, M reverses directly from $M_i$ to $M_f$ on an optimum ultra fast *quasi ballistic* trajectory. During ST pulse application M undergoes only a single precessional turn about $H_{int}$. At the moment of pulse decay M is oriented near $M_f$ and thus near the final reversed equilibrium orientation as sketched in (d) and the inset to (f). Now the complete reversal from the initial to the reversed equilibrium takes place on the time scale of only one precessional turn. This biased quasi ballistic ST reversal represents the fundamental ultra fast limit of the magnetization reversal time for the given field and current configuration.

As mentioned before *field induced* quasi ballistic reversal (25,25) can occur by a *half* precessional turn, only. During biased ST reversal M in principle passes near $M_f$ already after approximately a half precessional during the initial precessional overshoot observed in (c). However no reliable reversal has been found for the corresponding pulse duration. Most probably here M has a significant out of plane component corresponding to a large demagnetizing energy of the free layer. This strong excitation inhibits biased ballistic ST reversal by a half precessional turn.

Concluding we have presented room temperature measurements of time resolved ST precession and free ringing in MTJ nanopillars as used in present ST-MRAM prototypes. We have shown that coherent control schemes allow determining the final orientation of M upon decay of the ST pulse. Biased quasi ballistic ST magnetization reversal was demonstrated. Such optimized reversal could allow low current, ultra fast and highly reliable ST devices.

Acknowledgement: We acknowledge funding by the Deutsche Forschungsgemeinschaft within the Priority Programme "Ultra fast magnetization processes".

**FIGURE CAPTIONS:**

**Fig. 1:** (a) free layer easy axis TMR hysteresis loop. (b) free layer switching asteroid compiled from easy axis loops at different hard axis bias fields. Configuration of static field $H_S$, free M and pinned $M_P$ magnetization is sketched. (c) sketch of the experimental setup for electrical detection of ST precession.

**Fig. 2:** Time-resolved TMR signal of ST and free precession. $H_S = 22.8$ mT. (a) ST precession during the rise of a 5 ns pulse of j = $3.2 \times 10^6$ A/cm$^2$ at $\phi = -114°$. (c) free precession after excitation by a pulse of 180 ps of j = $9 \cdot 10^6$ A/cm$^2$ at $\phi = 84°$. (b) (d) show the precession data from (a),(c), respectively, after subtraction of an exponential background (red line in (a),(c)). Precession parameters are derived from fitting an exponentially damped sinusoid to the precession data (red line).

**Fig. 3:** (a),(b): Suppressed ringing of M by variation of ST pulse duration. $\phi = -104°$, $H_S = 24$ mT, j = $-7.8 \times 10^6$ A/cm$^2$. $T_{Pulse} = 450$ps, 600 ps, 770ps, 930 ps, 1100ps, 1250 ps from bottom to top. Graphs are offset for clarity. Gray lines mark onset and decay of the pulses. For the black curves ringing is coherently suppressed after an integer number (n = 2,3,4) of full precessional turns. For the red curves ringing is enhanced as the ST pulse terminates after n-1/2 precessional turns as sketched in (b). (c)-(f): Biased precessional ST reversal by pulses of j = $2.5 \, 10^6$ A/cm$^2$ with $\mu_0 H_S = 10$ mT along hard axis. (c) measured ST reversal by 5 ns pulse. (d) sketch of magnetic configurations and reversal trajectory. A strong ST precession (i) and relaxation to reversed state (ii) after the pulse is observed. (e) ST reversal by 730 ps leads to strong ringing (iii) and low switching reliability. (f) quasi ballistic ST reversal by 1.1 ns pulse. Pulse terminates in precession minimum of (c). Reversal occurs on an optimized trajectory within one full precessional turn.

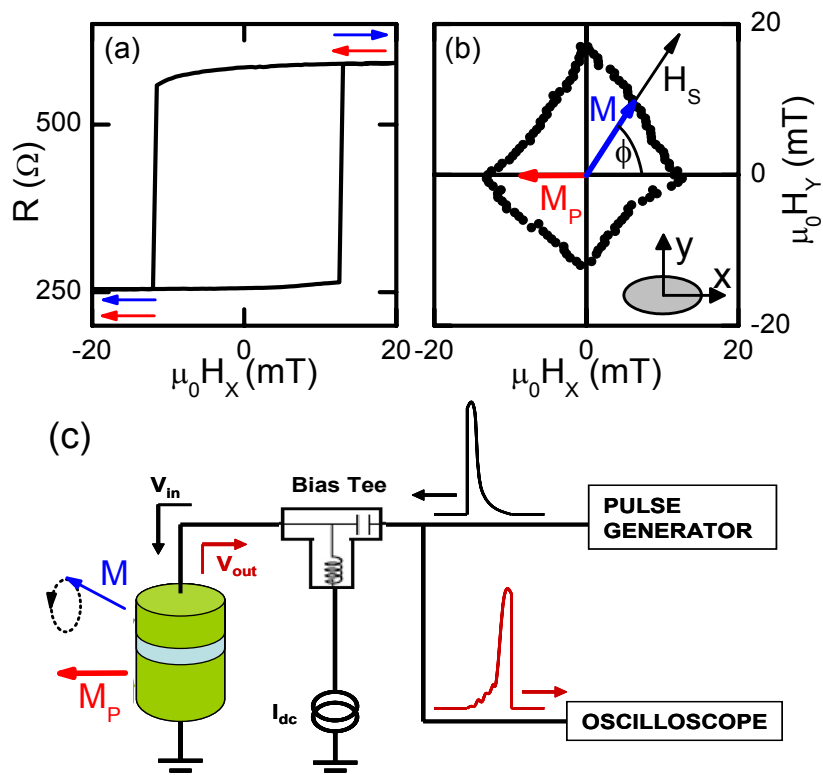

Figure 1
Serrano Guisan et al.

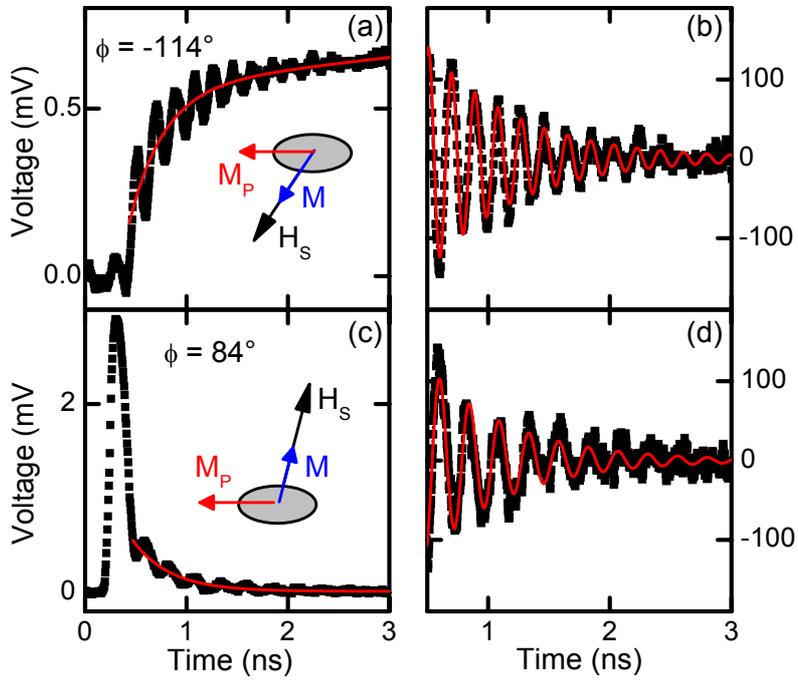

Figure 2
Serrano Guisan et al

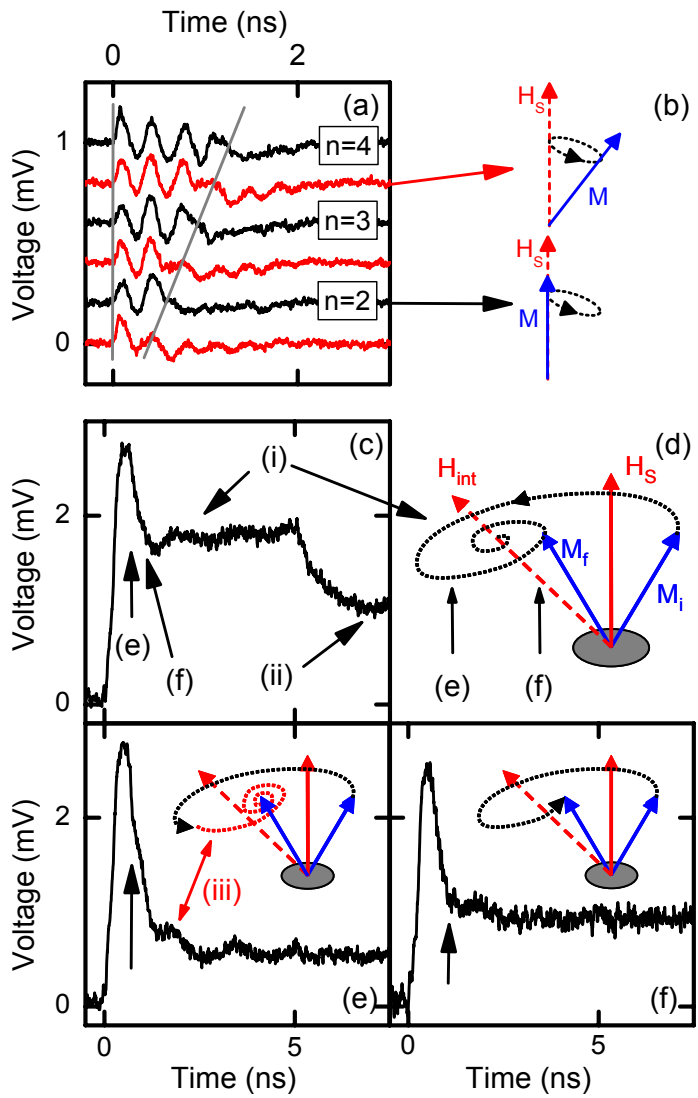

Figure 3
Serrano-Guisan et al.